# Flaw-Driven Failure in Nanostructures

*X. Wendy Gu[†], Zhaoxuan Wu[°], Yong-Wei Zhang[°], David J. Srolovitz[+], Julia R. Greer[‡]\**

[†]Division of Chemistry and Chemical Engineering, and [‡]Division of Engineering and Applied Science, California Institute of Technology, 1200 E. California Blvd., Pasadena, CA 91125, United States

[°]Institute of High Performance Computing, 1 Fusionopolis Way, #16-16 Connexis, Singapore 138632

[+]Departments of Materials Science and Engineering & Mechanical Engineering and Applied Mechanics, University of Pennsylvania, Philadelphia, PA 19104, United States

CORRESPONDING AUTHOR:

Julia R. Greer

M/C 309-81, Division of Engineering and Applied Science, California Institute of Technology, 1200 E. California Blvd., Pasadena, CA 91125, United States

626-395-4127

jrgreer@caltech.edu

**Abstract**

Understanding failure in nanomaterials is critical for the design of reliable structural materials and small-scale devices that have components or microstructural elements at the nanometer length scale. No consensus exists on the effect of flaws on fracture in bulk nanostructured materials or in nanostructures. Proposed theories include nanoscale flaw tolerance and maintaining macroscopic fracture relationships at the nanoscale with virtually no experimental support. We explore fracture mechanisms in nanomaterials via nanomechanical experiments on nanostructures with pre-fabricated surface flaws in combination with molecular dynamics simulations. Nanocrystalline Pt cylinders with diameters of ~120 nm with intentionally introduced surface notches were created using a template-assisted electroplating method and tested in uniaxial tension in *in-situ* SEM. Experiments demonstrate that 8 out of 12 samples failed at the notches and that tensile failure strengths were ~1.8 GPa regardless of whether failure occurred at or away from the flaw. These findings suggest that failure location was sensitive to the presence of flaws, while strength was flaw-insensitive. Molecular dynamics simulations support these observations and show that incipient plastic deformation commences via nucleation and motion of dislocations in concert with grain boundary sliding. We postulate that such local plasticity reduces stress concentration ahead of the flaw to levels comparable with the strengths of intrinsic microstructural features like grain boundary triple junctions, a phenomenon unique to nano–scale solids that contain an internal microstructural energy landscape. This mechanism causes failure to occur at the "weakest link," be it an internal inhomogeneity or a surface feature with a high local stress.

**Text**

Hard biomaterials such as shell, bone and exoskeletons have exceedingly high strength and fracture toughness that are on par with the best manmade structural materials[1,2]. These biomaterials have a unifying feature; their internal structures are hierarchically arranged, with distinct features on length scales extending from the nano to the macro. Nanofabrication techniques have advanced to the point where it is now possible to emulate these hierarchical structures, for example using ultra-high strength nanoscale building blocks made of carbon and inorganic nanotubes (1D) and platelets (2D), and metals with nanoscale interfaces (3D) as the load-bearing components[3-8]. The high intrinsic strength of these nanomaterials is often difficult to maintain in large-scale composites because a macroscopic ensemble of these structures routinely contain structural and/or chemical flaws within individual constituents or at the interfaces, which are sources of failure-initiation[9,10]. Classical fracture mechanics dictates that susceptibility to fracture depends on sample and/or external flaw length scales. This implies that different behavior may occur at small sample sizes and that new fracture relations may be necessary to describe failure of nanoscale materials[11,12].

Several theoretical and computational studies have been performed on fracture in pre-flawed nanoscale samples, often leading to conflicting interpretations. In the theoretical work of Gao *et al.*, scaling arguments based on linear elastic fracture mechanics (LEFM) were used to define a critical length, 0.2-400 nm for typical brittle materials, below which the strength of a hard platelet becomes comparable to the theoretical strength of the material regardless of the presence of structural flaws[13]. This nanoscale flaw tolerance, or flaw insensitivity, has been proposed as an explanation for the extraordinary toughness found in experiments on nanostructured biomaterials like nacre and spider silk[13,14] and in atomistic simulations of nanocrystalline aluminum thin films and polycrystalline graphene sheets, which showed failure occurring away from the pre-fabricated hole[15,16]. Other studies reported a strong dependence of failure on the presence of flaws – for example, in graphene and carbon nanotubes, where intentionally introduced holes led to strengths that are well below theoretical predictions, but in good agreement with predictions based upon classical fracture mechanics[17]. Even very small holes in a carbon nanotube sidewalls consisting of 1-6 missing atoms were shown to reduce the nanotube strength by as much as 26-33%[18].

Few well-controlled experimental fracture tests have been attempted at the nanoscale. An *in situ* transmission electron microscopy (TEM) study of a tensile-loaded nanocrystalline aluminum thin film containing a focus ion beam (FIB) milled edge notch (50 nm radius), showed failure occurring far from the notch[19]. Traditional fracture testing methodologies have been extended to the micron-scale using FIB milled cantilever coupons (1-10 μm in size) to study fracture in single and bi-crystalline metals and alloys[20-22]. Results demonstrate that these micron-sized metals fractured as predicted by LEFM, with fracture strength and location controlled by the FIB-milled structural flaw.

These studies prompt several important questions about fracture at the nanoscale, including 1) Does fracture strength depend on the presence of flaws and on sample/flaw geometry? and 2) Can the initiation point of the crack that leads to failure be predicted based upon the location of the flaw? We address these questions by conducting tensile fracture experiments and molecular dynamics (MD) simulations on similar nano-sized samples with introduced surface flaws of known geometries. Nanocrystalline Pt, referred to as nc-Pt hereafter, nanocylinders with surface notches were fabricated through pulsed electroplating into poly-methyl-methacrylate templates and do not suffer from FIB-induced damage common to many nanomechanical experiments. Typical Pt samples were ~120 nm in diameter with the grain size of ~6 nm. Tensile experiments on unnotched nc-Pt nanocylinders revealed brittle failure; rendering them to be an appropriate material model system for testing fracture mechanics theories that assume limited plasticity. Sizes of surface notches in these nanostructures were of the same order of magnitude as internal microstructural features, i.e., grain size. We examine the competition between such pre-fabricated surface and pre-existing microstructural flaws as preferred sites for crack initiation and discuss these findings using concepts from LEFM, weakest link theory and atomistic simulations.

Nanocrystalline platinum cylinders with diameters of 117 ± 3 nm, and lengths of 750 ± 40 nm were fabricated with one or more surface flaws using a template-assisted electroplating method described in Gu et al.[23]. The surface flaws generally had the shape of a rounded notch (See Fig.1a and b). The circumferential length of the notch, $b$, and height of the notch, $h$, were identified experimentally using scanning electron microscopy (SEM) imaging of the front and back faces of the nanocylinder. The notch depth, $a$, and radius, $r$, were estimated using this method, but cannot be determined precisely because SEM imaging cannot be performed at the necessary angles relative to the notch for complete characterization of $a$ and $r$. Notch geometries were grouped into two categories: (1) a straight notch or (2) a partial circumferential notch based on SEM images, with $r$ equal to half of $h$ (see Supporting Information, Fig. S1). Notch geometry was described in terms of the fraction of cylinder circumference, $\bar{b} = \frac{b}{\pi D}$, and fraction of the cylinder height, $\bar{h} = \frac{h}{l}$, for ease of comparison across experiment and simulation. Resulting unitless dimensions were $\bar{b}$=0.10-0.50 (circumferential length of $b$=40-200 nm) and $\bar{h}$=0.02-0.07 (notch height $h$=15-50 nm) (Fig. 1). Transmission electron microscopy (TEM) revealed the grain size to be 6 ± 3 nm with no significant variation across sample volumes. Size and shape of nanoscale samples and flaws play an important role in failure processes, hence to compare fracture properties in a nanomaterial with those with macroscopic dimensions, it is necessary to follow a standard for nanomechanical testing. Existing ASTM fracture standards are designed for large samples; no fracture testing standard exists for nanoscale samples[24]. This work serves as a step towards establishing this standard because it sheds light on fundamental physics of fracture mechanisms in nanomaterials in the presence of notches, which play a key role in failure of macroscopic samples. The sample geometry in this work is appropriate for nano-fracture testing because the surface flaws represent a major stress concentrator at which failure initiation would be expected in a typical macroscopic sample. The nc-Pt samples failed at ~3% strain with no observable bending at the flaw. Plastic zone size was estimated to be 30 nm, ~1/4 of sample diameter, using the critical stress intensity factor for an almond-shaped crack in a solid cylinder, a reasonable approximation of the actual notch geometry[20,25].

The samples were oriented during mechanical tests such that the surface notch was on the side (rather than the front or back faces) of the sample relative to the imaging electron beam to observe the initiation of failure (see Fig. 2). We found that 8 out of 12 samples broke at the surface flaw and the remaining 4 broke away from the flaw. Stress-strain data for each experiment showed brittle failure, with limited plastic deformation and no noticeable necking (Fig. 2 a,d) and with no significant difference in ultimate tensile strengths (UTS) between samples that broke at the flaw (UTS of $1.8 \pm 0.1$ GPa), ones that broke away from the flaw (UTS of $1.8 \pm 0.2$ GPa), and the unnotched samples. This strength is 50% higher than that of similarly-fabricated Pt nanopillars tested in compression, which is likely due to the higher deformation strain rate ($0.001$ s$^{-1}$ vs. $0.01$ s$^{-1}$ here) and the tension-compression asymmetry present in nanocylinders [23,26].

SEM images revealed features with dimensions on the order of grain size that populated the fracture surfaces, reminiscent of typical dimpled fracture morphology of bulk nanocrystalline metals[27,28] (see *SI Text*, Fig. S2). The angle of the fracture surface relative to the loading axis and the curvature of the fracture surface across the width of the broken cylinder were unpredictable.

The finding that 2/3 of the samples broke at the notch suggests the sensitivity of failure initiation to flaws. Both sets of cylinders – ones that failed at the flaw and those that failed away from it – exhibited nearly identical fracture strengths, which implies flaw-insensitivity in strength. To resolve this apparent contradiction, we performed molecular dynamics simulations of nc-Pt samples with notch and sample geometries similar to those in the experiments to reveal the mechanistic origin of the experimentally observed deformation and failure. We first created a polycrystalline simulation cell with an average grain size of ~14 nm, from which we carved out a notch-free cylinder of 43 nm diameter and 206 nm length. In addition to the notch-free nanocylinder (Fig. 3a), we used the same nanocylinder to create four additional samples, each containing a straight notch with a different configuration: (1) $\bar{b} = 0.16$ and $\bar{h} = 0.03$ (b = 21 nm and h = 5 nm) (Fig. 3b), (2) $\bar{b} = 0.2$ and $\bar{h} = 0.03$ (b = 30 nm and h = 5 nm) (Fig. 3c), (3) $\bar{b} = 0.23$ and $\bar{h} = 0.006$ (b = 31 nm and h = 1 nm) (Fig. 3d), and (4) $\bar{b} = 0.33$ and $\bar{h} = 0.006$ (b = 44 nm and h = 1 nm) (Fig.3e). Following equilibration at room temperature, the nanocylinders were uniaxially stretched to failure[29]. Before creating the notch, we first identified the fracture location in the notch-free nanocylinder, and then placed the notch far away from the fracture location.

Figure 3 shows the undeformed configurations (a1-e1), the deformed configurations (a2-e2), and the stress-strain data (a3-e3) for all five nanocylinders. Simulations revealed that samples in Fig. 3b and d were not affected by the pre-existing notches, in contrast to those in Fig. 3c and e, which failed at the pre-existing notches. Regardless of the location of failure initiation, all stress-strain curves are remarkably similar: they all exhibit nearly linear elastic behavior up to the UTS of ~3 GPa, followed by rapid strain softening. The nearly identical UTS in all five samples demonstrates that the UTS was insensitive to the presence of notches and the occurrence of failure both at and away from the notches suggests sensitivity of failure initiation to flaws, which corroborates the experiments.

The experimental and computational results present compelling evidence that the effects of notches on deformation and failure of nanomaterials are significantly different from those in their coarse-grained counterparts. To gain fundamental insight into what makes failure of nanomaterials different in response to external notches, we examined these processes at the atomic level. Figure 4 shows the spatial distribution of atomic-level (virial) tensile stresses ($\sigma_{yy}$) at an applied strain of 2.5%, as indicated in their respective stress-strain curves in Fig. 3. This analysis revealed that the initiation of failure was defined by the weakest link within the sample, whether it is near the notch or at an internal microstructural feature, and that failure always occurred via dislocation plasticity where the stress could not be relieved by grain boundary sliding alone.

Fig. 4a shows the internal stress concentrations at the grain boundaries and at the triple junctions within the notch-free nanocylinder. These stress concentrations are related to both the elastic anisotropy within the material and to grain boundary sliding[23]. The stress concentration factor at triple junctions reached 2-3 times the average stress, which is apparently large enough to trigger dislocation nucleation, as illustrated in Fig. 4a3. Subsequent deformation was highly localized and was characterized by grain boundary sliding with limited dislocation activity in grains near the surface[23]. In the interior, grain boundary sliding was more limited, presumably because of the constraint from other grains. Simulations revealed that the intrinsic failure mechanism of the nc-Pt nano-cylinders was strongly localized plasticity associated with grain boundary sliding accommodated by limited dislocation activity (see Supporting Information for movies of the fracture process).

Figure 4b shows a cylinder with a half-cylindrical notch. A close comparison of Fig. 4a and b shows that the stress distributions were nearly identical in the notched and unnotched samples. The stress concentration at the notch root in Fig. 4b was similar in magnitude but had a narrower range compared to those at many grain boundary triple junctions elsewhere in the cylinder. As the applied strain increased, dislocations were nucleated close to the notch root and propagated toward grain boundaries at the opposite side of the grain (see Fig. 4b3). Severe grain boundary sliding and dislocation motion were observed in the grains close to the notch root; this led to a reduction in the local stresses and to notch root blunting, which shielded this region from further stress increase (see Supporting Information for movies of stress evolution during fracture). These localized plastic events were imperceptible in the stress-strain curves. Subsequent to these localized events near the notch, the deformation of the notched and unnotched cylinders was very similar: incipient dislocation nucleation and propagation occurred at the microstructural features within the cylinder (away from the notch) with the highest stress concentrations. This explains the nearly identical fracture morphology in the notched vs. unnotched samples (see Fig. 3a2-b2) and suggests that the notch root was not necessarily the weakest link in all samples. The nature of failure was virtually unchanged in the presence or in the absence of a surface flaw.

Figure 4c shows the cylinder with a deep half-cylindrical notch. Compared to the notched cylinder in Fig. 4b, this sample has a stronger stress concentration at the notch root: larger in magnitude and extending further from the notch root. In addition, stresses at grain boundaries near this notch are higher (*cf.* Fig. 4 b2 and c2) than in the previous case. As in the previous case, plastic deformation close to the notch root led to notch root blunting, and this localized plasticity had little effect on the observed stress-strain response In addition to the weakest link location in the notch-free sample, plasticity in this notched sample also initiated at the notch root because of the higher stresses in the grains close to the notch. This suggests the presence of two weakest links, one external (Fig. 4c) and one internal (Fig. 4b), both undergoing substantial plastic deformation during loading. After reaching the ultimate strength, plastic deformation in this notched cylinder was more localized at the notch root, resulting in a different failure location from the unnotched sample and the sample with the shallower notch. Despite the change in fracture location with increasing notch size, the stress-strain curve is nearly identical with the unnotched and small notch samples.

Figure 4d shows a sample with an atomically sharp notch. Stress concentration was highest at the notch root and most localized amongst the cases in Fig. 4a-d. A comparison between Fig. 4c2 and d2 shows that the stress concentration in the grains near the notch in Fig. 4d2 was actually lower than that in Fig. 4c2. Similar to the sample shown in Fig. 4c, both the original weakest link location and the location close to the notch root exhibited substantial plastic deformation during loading (see Fig. 4d3). At a larger applied strain, the plastic deformation at the original weakest link became dominant and final failure occurred there. Hence, although the atomically sharp notch could generated a high stress concentration before the occurrence of plasticity at the notch root, it did not dictate the failure location for this nanocylinder. Close examination of the stress evolution in Fig. 4d shows that the local plasticity ahead of the notch root effectively reduced the original stress concentration, which shifted the weakest link from the sharp notch to the previous grain boundary triple junction.

Figure 4e shows a sample with a deep, atomically sharp notch. Here, the stress concentration at the notch tip is large in magnitude and extends far into the sample. The tensile stresses in the grains near the notch was substantially higher than those in the previous cases shown in Fig. 4b2-d2. During loading, extensive plastic deformation occurred only within the grains close to the notch root. This notch created a stress concentration which is larger in magnitude and in extent than at any internal inhomogeneity, which led to failure at the notch rather than at an internal microstructural feature and caused a less than 5% reduction in UTS as compared with the other cases. In this case, the notch served as the weakest link and dictated the location of failure.

The fundamental picture of failure in nanocrystalline nanostructures with surface flaws that emerges from experiments and simulations is that these materials fail at the location of the "weakest link" where the intrinsic failure criterion is first met. These weak links are well correlated with regions of high stress; nonetheless, the fundamental failure mechanisms are invariably the same, that is, failure occurs by localized grain boundary sliding and local accommodation via dislocation nucleation (usually at grain boundary triple junctions) and propagation across grains to be reabsorbed at grain boundaries. Since the fundamental failure mechanisms are the same in the presence or absence of notches, the UTS is insensitive to exactly where the failure process starts. The location of failure initiation can be associated with either external or internal stress concentration sources. The external sources of stress concentrations are surface flaws or notches, whose strength depends on the flaw size and sharpness, as well as on the local microstructure in the vicinity of the flaw tip. The microstructure enters as the source of dislocations that can blunt the tip and hence such relaxation gives rise to a stochastic element for identical flaws. The dominant internal stress concentrators in these polycrystalline samples are grain boundary triple junctions, which activate dislocation nucleation through grain boundary sliding. Such features are ubiquitous within nanocrystalline samples where the sample diameter is large compared to the grain size. The strength of triple junctions as stress concentrators depends on the orientation of the sliding grain boundaries relative to the load as well as on the orientation of the slip systems within the grains for easy dislocation nucleation. This gives rise to a statistical distribution of stress concentrators and hence to a distribution of weak links. Failure initiates at the weakest link in the system regardless of whether it is an internal or an external stress concentrator, which explains the stochastic nature of failure initiation location observed in experiments and in simulations. These phenomena are unique to nano-scale solids because their sample dimensions span tens of grains as opposed to thousands or greater as is the case in macroscopic samples. In a macroscale, nanocrystalline system, the UTS can be modulated by introduction of very large, sharp flaws that create much greater stress concentrations than the internal stress landscape.

Although the MD simulations demonstrate excellent qualitative agreements with experimental results, quantitative differences exist. MD simulations show a moderately ductile fracture on the scale of the grain size, while the experiments suggest a "brittle" fracture process. This discrepancy could be attributed to the differences in grain sizes (6 nm in experiments vs 14 nm in simulations), the number and orientation of the grains across the cylinder diameter and ahead of notches, and from the 10 orders of magnitude difference in strain rates. Nevertheless, the experiments and simulations unambiguously demonstrate that failure initiation in nanomaterials is determined by the weakest link, and that the UTS is insensitive to the physical origin of the weakest link.

The comparable stress concentrations associated with the notch and with the microstructural weak spots demonstrate a breakdown in the applicability of continuum elastic theory to describe failure in nanostructured materials. In continuum theory, the stress concentrations at the notch root should be much higher than those observed in the MD simulations, although elastic theory only applies to the early stages of deformation, before the onset of plasticity. The effect of stress concentration on failure was also examined in the experiments. The notches on 11 of the 12 experimentally tested nanocylinders were carefully characterized using SEM and imported into continuum finite element modeling (FEM) models, not accounting for nanocylinder microstructure (see *SI Text,* Fig. S3 and S4). The calculations revealed that the stress concentrations at the notches in the cylinders that broke at the notch (2 to 6.5) to be higher than those for notched samples that broke elsewhere (1.7 to 2) for all but one sample (see *SI Text*, Fig. S5). This indicates the tendency for failure to initiate in regions where the stress concentration is large. Experimentally obtained UTS were unrelated to such calculated stress concentrations. The finding that the experimentally (and simulation) obtained UTS is insensitive to notch size and shape demonstrates that fracture strength is likely governed by microstructural effects. The limited statistical set of samples studied here shows a slightly larger range in the variation of UTS in the 4 samples that broke away from the notch indicates a wide distribution of local stress inhomogeneities in the nc-Pt; the range of UTS in the samples that broke at the flaw was 33% narrower, likely because fracture strength is governed by the self-similar structural flaws.

The results presented here can be summarized as: 1) flaw-insensitivity in strength: strength does not depend on whether failure initiates at an external flaw or within the microstructure, and 2) flaw-sensitivity in fracture location: nc-Pt nanocylinders tend to break at the pre-fabricated flaw regardless of fracture strength provided that the flaw is sufficiently large/sharp. These observations can be explained by a weakest link theory: nanostructures break at the weakest link. The weakest link is a region where the intrinsic failure criterion is first met, be it a microstructural feature such as grain boundary triple junctions or a surface flaw, such as the surface notches introduced here. Uniquely at the nanoscale, even sharp structural flaws that extend across a significant fraction of the sample may not govern failure because their effective stress concentrations are comparable to stress concentrations associated with internal flaws. This holds true as long as the external flaws are not dramatically larger than the relevant microstructural length scales.

This weakest link perspective naturally leads to flaw-sensitivity in failure initiation location, because incipient deformation and subsequent failure occur at the position of the weakest link. The localized plasticity in the vicinity of stress concentrators (triple junctions or flaws) tends to reduce the initially present stress concentrations, which leads to a more effective competition of the multiple stress concentrators throughout the sample volume. This process results in similar fracture strengths for a wide range of flaw shapes and sizes, which is manifested as flaw-insensitivity in strength.

Major structural flaws do not reduce the strength of nanoscale and nanostructured materials, yet may still serve as sites of failure initiation if the intrinsic failure criterion is reached because of the high local stress compared to stresses at internal, microstructural features. The high strength intrinsic to many nanostructures can be maintained while increasing fracture toughness, or resistance to failure at flaws, through microstructural toughening mechanisms. These findings shed light on failure processes in nanomaterials which commonly show significant deviations from behavior expected from classical continuum theory, and provide a physical foundation for the weakest link concept in failure of nanostructures. The present results suggest that future nanofracture testing be performed with careful consideration of microstructural effects as well as the well-defined/characterized sample/notch geometries. Sample/notch geometries appropriate for application of classical continuum theories may not be accessible in nanostructures where internal/microstructural flaws also have important influence on failure.

**Methods**

Nanocrystalline Pt nanocylinders with and without surface flaws were created using template assisted pulsed electroplating[30]. Cylinders were electroplated into nanoscale pores in PMMA (polymethyl methacrylate) on top of a conductive gold surface according to conditions in Gu *et al.*[23]. The PMMA was subsequently removed to leave freestanding cylinders. Approximately 120 nm diameter cylinders with flaws randomly distributed on the cylinder circumference were formed by ramping voltage from 0 V to 0.6 V at 85 mV/s, pausing the electroplating process for approximately five minutes, replacing the electroplating bath, and then applying two more pulses at the same voltage and plating rate. Applying three electroplating pulses was appropriate for filling the PMMA pore, and forming a hemispherical "head" above the PMMA layer that can subsequently be used as a grip during tension testing. We postulate that this fabrication technique leads to surface flaws because the first electroplating pulse leads to the formation of a columnar cylinder with several grains exposed on the top surface of the cylinder. The second set of pulses leads to the nucleation of new grains at some but not all of the exposed grains on the top surface of some of the cylinders. The flaw is formed where nucleation fails to occur between sets of electroplating pulses.

The geometry of each cylinder and its surface flaws was examined using scanning electron microscopy (SEM) at a 52° tilt at 0° (in order to image the front face of the cylinder) and 180° (back face) rotation, and at 86° tilt at 0° rotation (front face). Tension tests were performed in the SEMentor, an *in-situ* SEM with an attached nanoindenter, using a custom-milled diamond tension grip[31]. Electroplated Pt cylinders show poor adhesion to the underlying Au substrate, so a small amount of W glue was applied to the base of the cylinder using the FEI Nova 200 dual beam system. Tension tests were conducted at a constant strain rate of $0.01$ s$^{-1}$. SEM video was taken during tension testing and instrument compliances, changes in sample dimensions and fracture locations were determined from the video. Measured load-displacement data was converted to true stress-strain curves, after accounting for instrument compliance.

Sample preparation for TEM was performed by "plucking" a tension sample with the SEMentor[32]. To do this, the tension sample was fed into the SEMentor tension grip, which is used to lift the sample off the growth substrate. The grips were in contact with the sample on the underside of the tension head. The sample was then gently lowered onto a TEM grid using the tension grips, and then the tension grip was detached from the tension head. Carbon is applied to the base of the sample using e-beam deposition in order to glue the sample to the TEM grid.

Nanocrystalline nanocylinders were prepared for the molecular dynamics (MD) simulation by first creating a periodic simulation cell ($64 \times 206 \times 64$ nm) with 648 randomly placed "seeds" based upon which a Voronoi tessellation is performed. The resulting Voronoi polyhedra were then filled with atoms in a perfect face-centered cubic Pt crystal of random orientation to produce a nanocrystalline structure with an average grain size of ~14 nm. Nanocrystalline Pt nanocylinders of diameter ~43nm were carved from this periodic bulk nanostructure. MD simulations were performed using the Large-scale Atomic/Molecular Massively Parallel Simulator (LAMMPS), where interactions between Pt atoms were described using the Embedded Atom Method (EAM) potential parametrized by Sheng *et al.*[33-35]. Periodic boundary conditions were imposed along the nanocylinder axes, while surfaces of the nanocylinders were free. All of the nanocylinders were equilibrated at 300 K before tensile loading was applied[29]. The uniaxial tensile loading was applied by stretching the nanocylinders in the axial direction at a constant true strain rate of $0.1 \text{ ns}^{-1}$. During tensile loading, constant temperature was maintained using a Nosé- Hoover thermostat[36-39]. The atomic stresses were calculated based on the atomic *virial* stress in which the atomic volume was set to the Voronoi volume associated with each atom.

**Acknowledgements**

X.W.G. is grateful for financial support from the National Defense Science and Engineering Graduate (NDSEG) Fellowship, 32 CFR 168a. JRG acknowledges the financial support of the National Science Foundation (DMR-1204864). X.W.G. and J.R.G. thank the Kavli Nanoscience Institute at Caltech for the availability of critical cleanroom facilities. We thank V. Deshpande and D. Jang for helpful discussion, and D. Jang and C. Garland for TEM assistance. The authors gratefully acknowledge the financial support from the Agency for Science, Technology and Research (A*STAR), Singapore and the use of computing resources at the A*STAR Computational Resource Centre, Singapore.

# References

1. Currey, J. D. Mechanical-Properties of Mother of Pearl in Tension. *Proceedings of the Royal Society B-Biological Sciences* **196**, 443-463 (1977).
2. Fratzl, P. & Weinkamer, R. Nature's hierarchical materials. *Prog. Mater. Sci.* **52**, 1263-1334 (2007).
3. Yu, M. F. L., O.; Dyer, M. J.; Moloni, K.; Kelly, T. F.; Ruoff, R. S. Strength and breaking mechanism of multiwalled carbon nanotubes under tensile load. *Science* **287**, 637-640 (2000).
4. Lee, C., Wei, X. D., Kysar, J. W. & Hone, J. Measurement of the elastic properties and intrinsic strength of monolayer graphene. *Science* **321**, 385-388 (2008).
5. Garel, J. L., I.; Zhi, C. Y.; Nagapriya, K. S.; Popovitz-Biro, R.; Golberg, D.; Bando, Y.; Hod, O.; Joselevich, E. Ultrahigh Torsional Stiffness and Strength of Boron Nitride Nanotubes. *Nano Lett* **12**, 6347-6352 (2012).
6. Bertolazzi, S., Brivio, J. & Kis, A. Stretching and Breaking of Ultrathin MoS2. *ACS Nano* **5**, 9703-9709 (2011).
7. Lu, L., Chen, X., Huang, X. & Lu, K. Revealing the Maximum Strength in Nanotwinned Copper. *Science* **323**, 607-610 (2009).
8. Uchic, M. D., Dimiduk, D. M., Florando, J. N. & Nix, W. D. Sample dimensions influence strength and crystal plasticity. *Science* **305**, 986-989 (2004).
9. Vigolo, B. *et al.* Macroscopic fibers and ribbons of oriented carbon nanotubes. *Science* **290**, 1331-1334 (2000).
10. Hao, S. *et al.* A transforming metal nanocomposite with large elastic strain, low modulus, and high strength. *Science (New York, N.Y.)* **339**, 1191-1194 (2013).
11. Dugdale, D. S. Yielding of Steel Sheets Containing Slits *J. Mech. Phys. Solids* **8**, 100-104 (1960).
12. Hertzberg, R. W. *Deformation and Fracture Mechanics of Engineering Materials*. (Wiley, 1995).
13. Gao, H. J., Ji, B. H., Jager, I. L., Arzt, E. & Fratzl, P. Materials become insensitive to flaws at nanoscale: Lessons from nature. *Proc. Natl. Acad. Sci. U. S. A.* **100**, 5597-5600 (2003).
14. Giesa, T., Pugno, N. M. & Buehler, M. J. Natural stiffening increases flaw tolerance of biological fibers. *Physical Review E* **86**, 041902 (2012).
15. Kumar, S., Li, X. Y., Haque, A. & Gao, H. J. Is Stress Concentration Relevant for Nanocrystalline Metals? *Nano Lett* **11**, 2510-2516 (2011).
16. Zhang, T., Li, X. Y., Kadkhodaei, S. & Gao, H. J. Flaw Insensitive Fracture in Nanocrystalline Graphene. *Nano Lett* **12**, 4605-4610 (2012).
17. Khare, R. M., S. L.; Paci, J. T.; Zhang, S. L.; Ballarini, R.; Schatz, G. C.; Belytschko, T. Coupled quantum mechanical/molecular mechanical modeling of the fracture of defective carbon nanotubes and graphene sheets. *Phys. Rev. B* **75**, 075412 (2007).
18. Mielke, S. L. *et al.* The role of vacancy defects and holes in the fracture of carbon nanotubes. *Chem. Phys. Lett.* **390**, 413-420 (2004).
19. Kumar, S., Haque, M. A. & Gao, H. Notch insensitive fracture in nanoscale thin films. *Appl. Phys. Lett.* **94**, 253104 (2009).

20   Wurster, S., Motz, C. & Pippan, R. Characterization of the fracture toughness of micro-sized tungsten single crystal notched specimens. *Philosophical Magazine* **92**, 1803-1825 (2012).
21   Iqbal, F., Ast, J., Goken, M. & Durst, K. In situ micro-cantilever tests to study fracture properties of NiAl single crystals. *Acta Mater.* **60**, 1193-1200 (2012).
22   Kupka, D. & Lilleodden, E. T. Mechanical Testing of Solid-Solid Interfaces at the Microscale. *Experimental Mechanics* **52**, 649-658 (2012).
23   Gu, X. W., Loynachan, C. N., Wu, Z. X., Zhang, Y. W., Srolovitz, D. J., Greer, J. R. Size-Dependent Deformation of Nanocrystalline Pt Nanopillars. *Nano Lett* **12**, 6385-6392 (2012).
24   Wilson, C. D. & Landes, J. D. *Fracture toughness testing with notched round bars*. Vol. 30 (ASTM STP 1360, 2000).
25   Liu, A. *Summary of Stress-Intensity Factors*. Vol. 19 (ASM International, 1996).
26   Jang, D. C. & Greer, J. R. Size-induced weakening and grain boundary-assisted deformation in 60 nm grained Ni nanopillars. *Scr. Mater.* **64**, 77-80 (2011).
27   Dalla Torre, F., Van Swygenhoven, H. & Victoria, M. Nanocrystalline electrodeposited Ni: microstructure and tensile properties. *Acta Mater.* **50**, 3957-3970 (2002).
28   Hasnaoui, A., Van Swygenhoven, H. & Derlet, P. M. Dimples on nanocrystalline fracture surfaces as evidence for shear plane formation. *Science* **300**, 1550-1552 (2003).
29   Wu, Z. X., Zhang, Y. W., Jhon, M. H., Gao, H. J. & Srolovitz, D. J. Nanowire Failure: Long = Brittle and Short = Ductile. *Nano Lett* **12**, 910-914 (2012).
30   Burek, M. J. & Greer, J. R. Fabrication and Microstructure Control of Nanoscale Mechanical Testing Specimens via Electron Beam Lithography and Electroplating. *Nano Lett* **10**, 69-76 (2010).
31   Kim, J. Y., Jang, D. C. & Greer, J. R. Insight into the deformation behavior of niobium single crystals under uniaxial compression and tension at the nanoscale. *Scr. Mater.* **61**, 300-303 (2009).
32   Jang, D., Li, X., Gao, H. & Greer, J. R. Deformation mechanisms in nanotwinned metal nanopillars. *Nat Nanotechnol* **7**, 594-601, doi:10.1038/nnano.2012.116 (2012).
33   Plimpton, S. Fast Parallel Algorithms for Short-Range Molecular-Dynamics *J. Comput. Phys.* **117**, 1-19 (1995).
34   Daw, M. S. & Baskes, M. I. Embedded-Atom Method - Derivation and Application to Impurities, Surfaces, and Other Defects in Metals. *Phys. Rev. B* **29**, 6443-6453 (1984).
35   Sheng, H. W., Kramer, M. J., Cadien, A., Fujita, T. & Chen, M. W. Highly optimized embedded-atom-method potentials for fourteen fcc metals. *Phys. Rev. B* **83**, 134118 (2011).
36   Nose, S. A Unified Formulation of the Constant Temperature Molecular-Dynamics Methods. *J. Chem. Phys.* **81**, 511-519 (1984).
37   Nose, S. A Molecular-Dynamics Method for Simulations in the Canonical Ensemble *Mol. Phys.* **52**, 255-268 (1984).
38   Hoover, W. G. Constant-Pressure Equations of Motion *Phys. Rev. A* **34**, 2499-2500 (1986).
39   Melchionna, S., Ciccotti, G. & Holian, B. L. Hoover NPT Dynamics for Systems Varying in Shape and Size *Mol. Phys.* **78**, 533-544 (1993).
40   Kelchner, C. L., Plimpton, S. J. & Hamilton, J. C. Dislocation nucleation and defect structure during surface indentation. *Phys. Rev. B* **58**, 11085-11088 (1998).

**Figures**

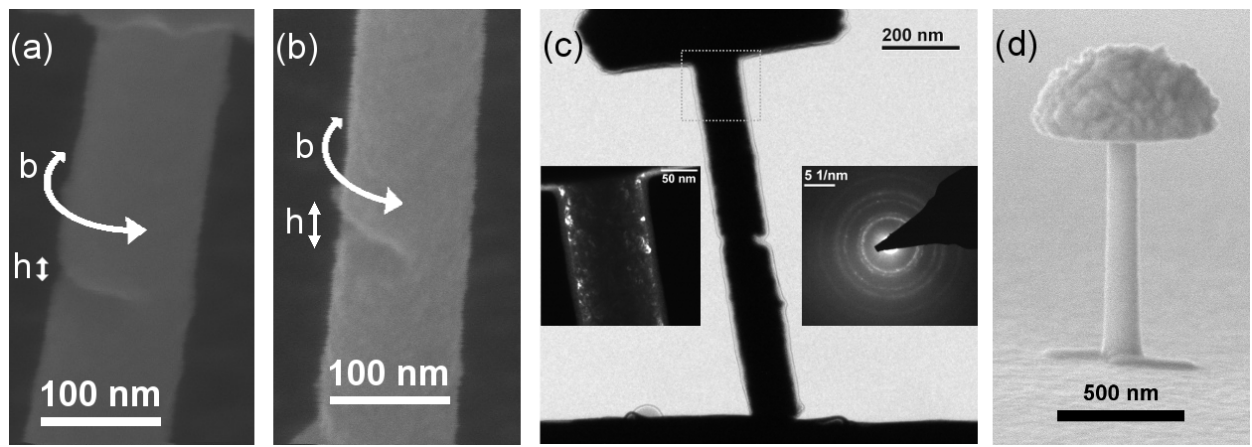

Figure 1. Nanocrystalline nanocylinders with intentionally introduced notches. SEM images taken from a 52° tilt of (a) a notch with circumferential length b = 84 nm, and height h = 47 nm ($\bar{b}$ = 0.23 and $\bar{h}$ = 0.06 when normalized by sample dimensions), and (b) a notch of b = 161 nm and h=24 nm ($\bar{b}$ = 0.54 and $\bar{h}$ = 0.03). (c) Bright-field TEM image of plucked cylinder, with boxed region represented in the dark-field image inset which shows nanocrystalline microstructure. The other inset shows the corresponding diffraction pattern. (d) SEM image of an un-flawed cylinder.

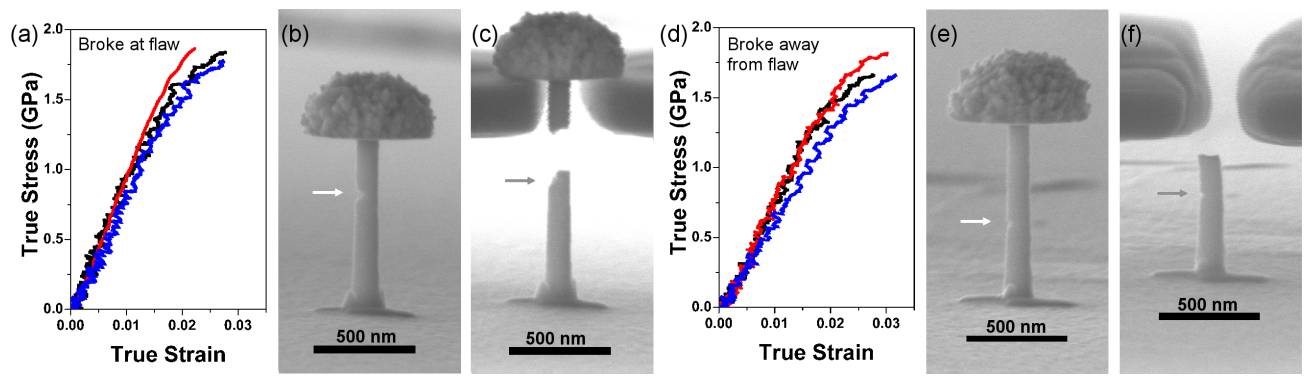

Figure 2. Samples that broke at the flaw: (a) representative true stress-true strain plots from uniaxial tension tests, (b) SEM image of a pre-flawed sample, and (c) SEM image of the same sample after fracturing at the flaw. Samples that broke away from the flaw: (d) representative true stress-true strain plots from uniaxial tension tests, (e) SEM image of a pre-flawed sample, and (f) SEM image of the same sample after fracturing away from the flaw.

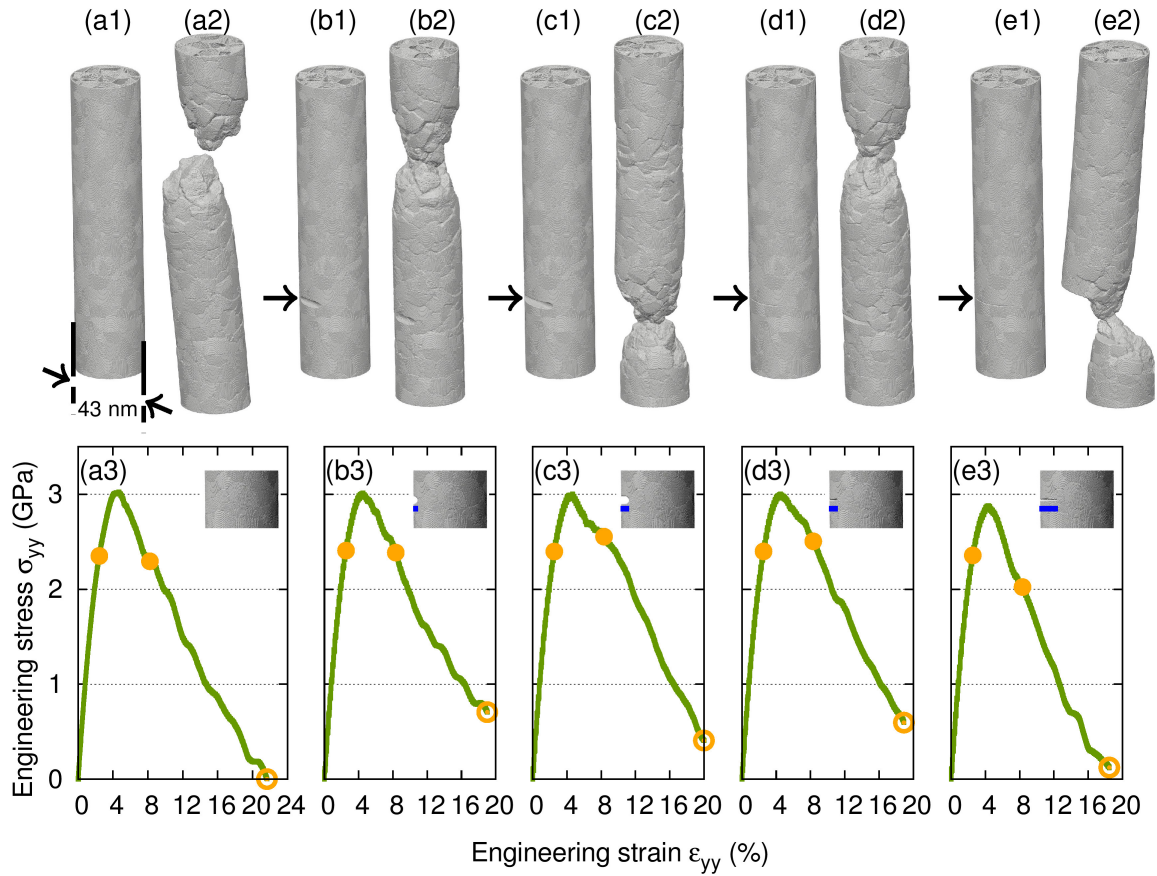

Figure 3. Undeformed (a1-e1) and deformed (a2-e2) simulation samples and corresponding stress-strain plots (a3-e3) for (a) notch-free, and notched samples with notch geometry with normalized circumferential width and height (b) $\bar{b} = 0.16$ and $\bar{h} = 0.03$ (b = 21 nm and h = 5 nm), (c) $\bar{b} = 0.2$ and $\bar{h} = 0.03$ (b = 30 nm and h = 5 nm), (d) $\bar{b} = 0.23$ and $\bar{h} = 0.006$ (b = 31 nm and h = 1 nm), and (e) $\bar{b} = 0.33$ and $\bar{h} = 0.006$ (b = 44 nm and h = 1 nm). In the stress-strain curves, the filled circles mark the 2.5% strain at which the atomic stresses in Fig. 2 were measured. The open circles mark the strains corresponding to the respective nanostructures shown above. The insets in (a3-e3) show the local views near the notches and the blue lines indicate the notch depths.

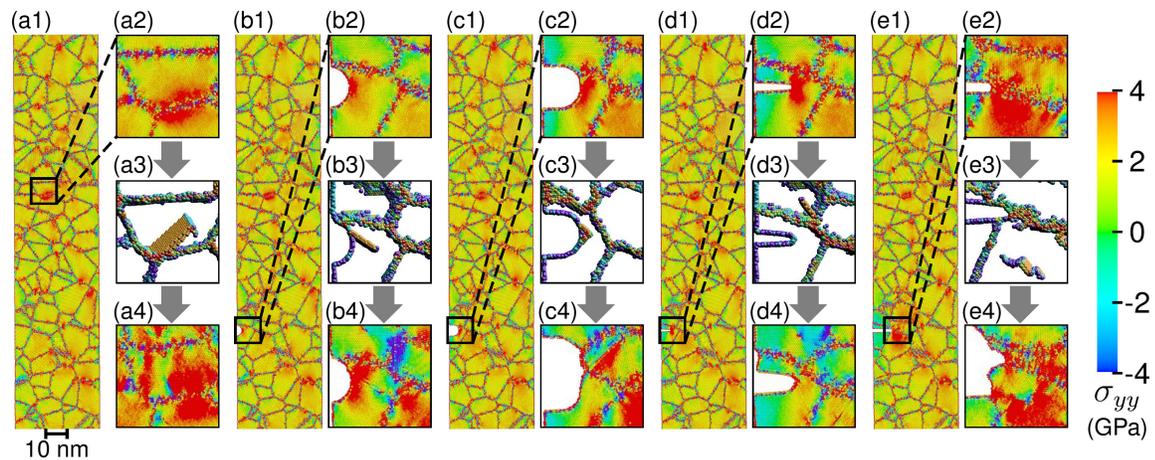

Figure 4. Cross-sectional view of the tensile stress ($\sigma_{yy}$) at 2.5% applied strain at (a1-e1) the mid-section of the samples shown in Fig. 3, and (a2-e2) magnified at a grain boundary triple junction and/or the notch root in order to highlight the stress concentration at these locations. (a3-e3) show the subsequent dislocation plasticity in the magnified view. Atoms are shown only if their central symmetry parameters differ from that of the perfect FCC crystal; the colours indicate the local symmetry [40]. Atoms on twin boundaries, dislocations, intrinsic and extrinsic stacking faults are shown in light blue, dark blue or green (depending on dislocation type), orange and light blue, respectively. (a4-e4) show the stress contours close to the previously identified high stress concentration points at 8% applied strain. Substantial stress reductions are apparent at notch roots (c4-e4) and grain boundary triple junction (a4), while (b4) shows a strong back stress arising from a dislocation ahead of the notch.